\documentclass[10pt,twocolumn,twoside] {IEEEtran}
\usepackage{booktabs}
\usepackage{multirow}
\usepackage{bm}
\usepackage{array}
\usepackage{graphicx}
\usepackage{epstopdf}
\usepackage{subfigure}
\usepackage{amssymb}
\usepackage{amsmath}
\usepackage{makecell}
\usepackage{bm}

\def\0{{\bf 0}}
\def\1{{\bf 1}}

\def\eg{{\em e.g.}}
\def\ie{{\em i.e.}}
\def\etc{{\em etc.}}

\usepackage[pagebackref=false,breaklinks=true,letterpaper=true,colorlinks,bookmarks=false]{hyperref}
\usepackage{cite}
\usepackage[linesnumbered,ruled,vlined]{algorithm2e}
\usepackage{multirow}
\usepackage{algorithmicx}

\def\eg{{\em e.g.}}
\def\ie{{\em i.e.}}
\def\etc{{\em etc.}}
\emergencystretch=\hsize
\graphicspath{{figs/}}

\begin{document}
%\linenumbers

\title{Adaptive Feature Selection Guided Deep Forest for COVID-19 Classification with Chest CT}

\author{Liang Sun$^\dagger$,
        Zhanhao Mo$^\dagger$,
        Fuhua Yan$^\dagger$,
        Liming Xia$^\dagger$,
        Fei Shan$^\dagger$,
        Zhongxiang Ding$^\dagger$,
        Wei Shao$^\dagger$,
        Feng Shi,
        Huan Yuan,
        Huiting Jiang,
        Dijia Wu,
        Ying Wei,
        Yaozong Gao,
        Wanchun Gao,
        He Sui,
        Daoqiang Zhang$^*$,
        Dinggang Shen$^*$,~\IEEEmembership{IEEE Fellow}

\thanks{L. Sun, W. Shao and D. Zhang are with the College of Computer Science and Technology, Nanjing University of Aeronautics and Astronautics, MIIT Key Laboratory of Pattern Analysis and Machine Intelligence, Nanjing 211106, China.

Z. Mo and H. Sui are with the Department of Radiology, The Third Hospital of Jilin University, Changchun, China.

F. Yan is with Department of Radiology, Ruijin Hospital, Shanghai Jiao Tong University School of Medicine, Shanghai, China.

L. Xia is with Department of Radiology, Tongji Hospital, Tongji Medical College, Huazhong University of Science and Technology, Wuhan, Hubei, China.

F. Shan is with Department of Radiology, Shanghai Public Health Clinical Center, Fudan University, Shanghai, China.

Z. Ding is with the Department of Radiology, Hangzhou First People¡¯s Hospital, Zhejiang University School of Medicine, Hangzhou, Zhejiang, China.

F. Shi, H. Yuan, H. Jiang, D. Wu, Y. Wei, Y. Gao and D. Shen are with the Department of Research and Development, Shanghai United Imaging Intelligence Co., Ltd., Shanghai, China

W. Gao is with the Department of Radiology, Qianjiang Central Hospital, Jishou University School of Medicine, Chongqing, China.}
\thanks{$^\dagger$: Equal contribution}
\thanks{$^*$Corresponding authors: D. Zhang (dqzhang@nuaa.edu.cn) and D.Shen (dinggang.shen@gmail.com).}
\thanks{This work was supported by the National Key Research and Development Program of China (Nos. 2018YFC2001600, 2018YFC2001602) and the National Natural Science Foundation of China (Nos. 61876082, 61861130366, 61732006, 61902183 and 81871337), the Royal Society-Academy of Medical Sciences Newton Advanced Fellowship (No.~NAF$\backslash$R1$\backslash$180371), and China Postdoctoral Science Foundation funded project (No. 2019M661831), Hubei Provincial Novel Pneumonia Emergency Science and Technology Project (No. 2020FCA021), the Novel Coronavirus Special Research Foundation of the Shanghai Municipal Science and Technology Commission (No. 20441900600).}
}

% The paper headers
\markboth{}%
{Sun~\MakeLowercase{\textit{et al.}}: Adaptive Feature Selection Guided Deep Forest}
% make the title area
\maketitle

\begin{abstract}
Chest computed tomography (CT) becomes an effective tool to assist the diagnosis of coronavirus disease-19 (COVID-19). Due to the outbreak of COVID-19 worldwide, using the computed-aided diagnosis technique for COVID-19 classification based on CT images could largely alleviate the burden of clinicians. In this paper, we propose an \textbf{A}daptive \textbf{F}eature \textbf{S}election guided \textbf{D}eep \textbf{F}orest (AFS-DF) for COVID-19 classification based on chest CT images. Specifically, we first extract location-specific features from CT images. Then, in order to capture the high-level representation of these features with the relatively small-scale data, we leverage a deep forest model to learn high-level representation of the features. Moreover, we propose a feature selection method based on the trained deep forest model to reduce the redundancy of features, where the feature selection could be adaptively incorporated with the COVID-19 classification model. We evaluated our proposed AFS-DF on COVID-19 dataset with 1495 patients of COVID-19 and 1027 patients of community acquired pneumonia (CAP). The accuracy (ACC), sensitivity (SEN), specificity (SPE) and AUC achieved by our method are 91.79\%, 93.05\%, 89.95\% and 96.35\%, respectively. Experimental results on the COVID-19 dataset suggest that the proposed AFS-DF achieves superior performance in COVID-19 vs. CAP classification, compared with 4 widely used machine learning methods.
\end{abstract}


%%%%%%%%%%%%%%%%% ^_^ ----------------- ^_^ %%%%%%%%%%%%%%%%%
\section{Introduction}
\IEEEPARstart{S}{ince} Decemeber 2019, the outbreak of coronavirus disease-19 (COVID-19)~\cite{zu2020coronavirus,choe2020coronavirus} has infected more than three million people worldwide, and causing more than 230, 000 deaths. World Health Organization (WHO) has declared the COVID-19 as a global health emergency on January 30, 2020~\cite{sohrabi2020world}. The chest computed tomography (CT) has shown to be useful to assist clinical diagnosis of COVID-19\cite{wu2020chest,fang2020sensitivity,li2020coronavirus,chung2020ct,liMingzhi2020coronavirus,long2020diagnosis}. However, the rapid growth of COVID-19 patients results in the shortage of the clinicians and radiologists. It is highly desired to develop automatic methods for computer-aided COVID-19 classification tools with chest CT images.

A few machine learning methods have been proposed for COVID-19 classification using chest CT images. For example,~\cite{shi2020deep} employs a logistic regression method for COVID-19 classification by using clinical and laboratory features.~\cite{tang2020severity,shi2020large} use random forest model with the handcrafted features for COVID-19 classification. Moreover, some deep learning based methods are proposed for the diagnosis of COVID-19. For instance, ~\cite{wang2020deep} leverages a deep learning method to learn the feature representation of the chest CT images, and then uses the learned features for COVID-19 classification by combining decision tree and Adaboost algorithm. In addition, ~\cite{xu2020deep} employs an end-to-end network to map the CT images to label space for COVID-19 disease identification.

In summary, the existing machine learning methods for COVID-19 diagnosis are mainly based on the handcrafted features or the learned image representation by deep neural networks. However, simply adopting handcrafted features cannot fully utilize the high-level information for COVID-19 classification, while the features learned from neural networks require great effort for parameter tuning with a small amount of medical image data.

\begin{figure*}[!tp]
			\setlength{\abovecaptionskip}{-2pt}
			\setlength\abovedisplayskip{-10pt}
			\setlength\belowdisplayskip{-10pt}
			\setlength{\belowcaptionskip}{-5pt}
			\centering
			\includegraphics[width=0.98\textwidth]{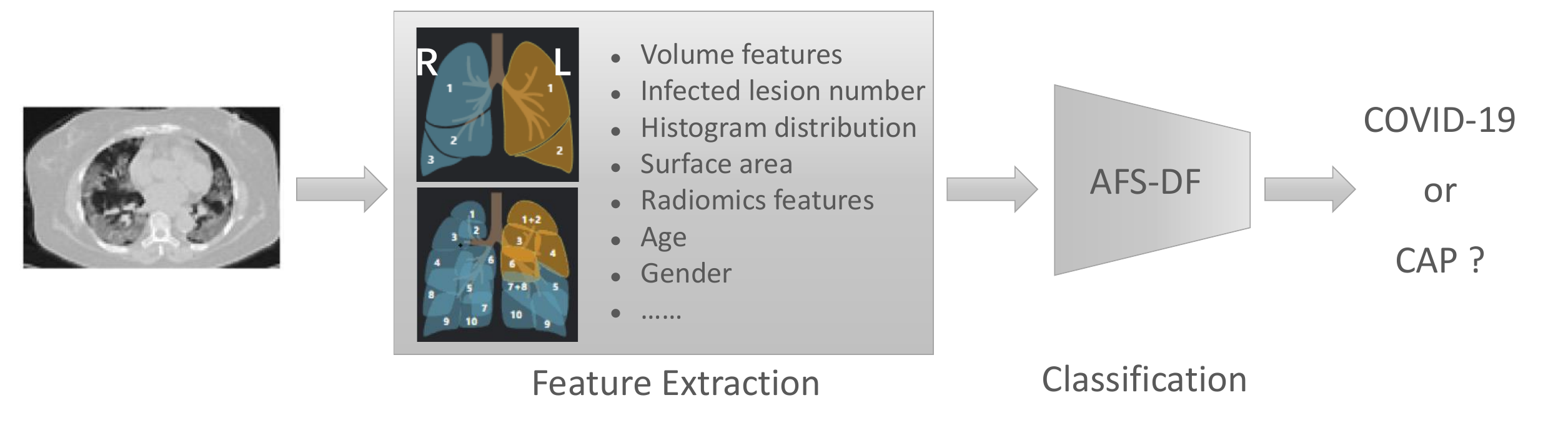}
			\caption{Pipeline of the proposed adaptive feature selection guided deep forest for COVID-19 vs. CAP classification. We first extract the location-specific features from chest CT. Then, the proposed AFS-DF is leveraged to train the classifier based on the location-specific features. Finally, based on the location-specific features, the trained FAS-DF is adopted for COVID-19 identification task. }
\label{fig1}
\end{figure*}

To this end, in this paper, we propose a novel adaptive feature selection guided deep forest method that takes advantage of the high-level deep features with small number of medical image data for the classification between COVID-19 and CAP. Specifically, as shown in Fig.~\ref{fig1}, we first extract the location-specific features from the chest CT image. Then, a deep forest model\cite{ijcai2017-497} is introduced to learn the latent high-level representations of these features, which can effectively describe the high-level information within the extracted location-specific features by using a small-scale training data. Intuitively, the use of the feature selection could promote the performance of the classification task. Hence, in our study, we also introduce a task-driven feature selection method to adaptively reduce the redundancy of features. In particular, the feature selection operation will discard a portion of unimportant features based on the feature importance which is calculated from trained forests in each deep forest layer. Hence, the feature selection and classifier training are adaptively incorporated into a unified framework. Finally, the trained adaptive feature selection guided deep forest is used for COVID-19 prediction.

The major contributions of this paper are three-fold. \emph{First}, a deep forest is leveraged to learn the high-level feature representation of the location-specific features from chest CT for the diagnosis of COVID-19. \emph{Second}, we proposed an adaptive feature selection method to adaptively select the discriminative features for the diagnosis of COVID-19. \emph{Third}, the proposed method is evaluated on the collected COVID-19 dataset, which consists of 1495 COVID-19 patients and 1027 CAP patients. Experimental results demonstrate that our method achieves superior classification performance than the comparison methods.

The rest of the paper is organized as follows. We first introduce the materials used in this study and the proposed adaptive feature selection guided deep forest in Section~\ref{Method}. Then, in Section~\ref{Experiment}, we present experimental settings and experimental results. In Section~\ref{Discussion}, we study the influence of parameters in the proposed methods and present the limitations of the current study as well as possible future directions. We finally conclude this paper in Section~\ref{Conclusion}.

%%%%%%%%%%%%%%%%------  New Section  -------%%%%%%%%%%%%%
\section{Materials and Method}
\label{Method}
In this section, we first introduce the dataset used in this study. Then, we present the feature extraction procedure for the location-specific features. Next, we describe the proposed adaptive feature selection guided deep forest (AFS-DF). Finally, we provide the implementation details for our proposed AFS-DF.

%%%%%%%%%%%%%%%%% ^_^ ----------------- ^_^ %%%%%%%%%%%%%%%%%
\subsection{Materials}
\label{Materials}
A total of 2522 chest CT images are used in our study, provided by The Third Hospital of Jilin University, Ruijin Hospital of Shanghai Jiao Tong University, Tongji Hospital of Huazhong University of Science and Technology, Shanghai Public Health Clinical Center of Fudan University, and Hangzhou First People¡¯s Hospital of Zhejiang University. In this dataset, 1495 cases are from the confirmed COVID-19 cases diagnosed by positive nucleic acid testing. The other 1027 case are from CAP patients. COVID-19 images are acquired from Jan. 9, 2020 to Feb. 14, 2020, and CAP images are obtained from Jul. 30, 2018 to Feb. 22, 2020. The demographic information of these 2522 subjects is summarized in Table~\ref{tab1}.

All patients underwent chest CT scans with thin section. Specifically, CT scanners include uCT 780 from UIH, Optima CT520, Discovery CT750, LightSpeed 16 from GE, Aquilion ONE from Toshiba, SOMATOM Force from Siemens, and SCENARIA from Hitachi. CT protocol includes: 120 $kV$, reconstructed slice thickness ranging from 0.625 to 2 $mm$, with breath hold at full inspiration. All images are de-identified before sending for analysis. The study are approved by the Institutional Review Board of participating institutes. Written informed consent is waived due to the retrospective nature of the study.

\begin{table}[tbp]
	\setlength{\belowcaptionskip}{-1pt}
	\setlength{\abovecaptionskip}{-1pt}
	\renewcommand\arraystretch{1.4}
	\centering
	\footnotesize
	\caption{Demographic information of the studied 2685 chest CT scans from COVID-19 dataset. M/F: Male/Female}
	\begin{tabular*}{0.48\textwidth}{@{\extracolsep{\fill}}ccc}
		\toprule
	    \multicolumn{1}{l}{Category} & Scan \#& Gender (M/F)\\
		\hline
        \multicolumn{1}{l}{\multirow{1}{*}{COVID-19}}
		&$1495$		 &$770/725$\\
		%\hline  		
		\multicolumn{1}{l}{  \multirow{1}{*}{CAP}}
		&$1027$		 &$488/539$\\
		\bottomrule
	\end{tabular*}
	\label{tab1}
\end{table}

%%%%%%%%%%%%%%%%% ^_^ ----------------- ^_^ %%%%%%%%%%%%%%%%%
\subsection{Feature Extraction}
\label{FE}
\begin{figure*}[!tp]
			\setlength{\abovecaptionskip}{-5pt}
			\setlength\abovedisplayskip{-10pt}
			\setlength\belowdisplayskip{-10pt}
			\setlength{\belowcaptionskip}{-5pt}
			\centering
			\includegraphics[width=0.98\textwidth]{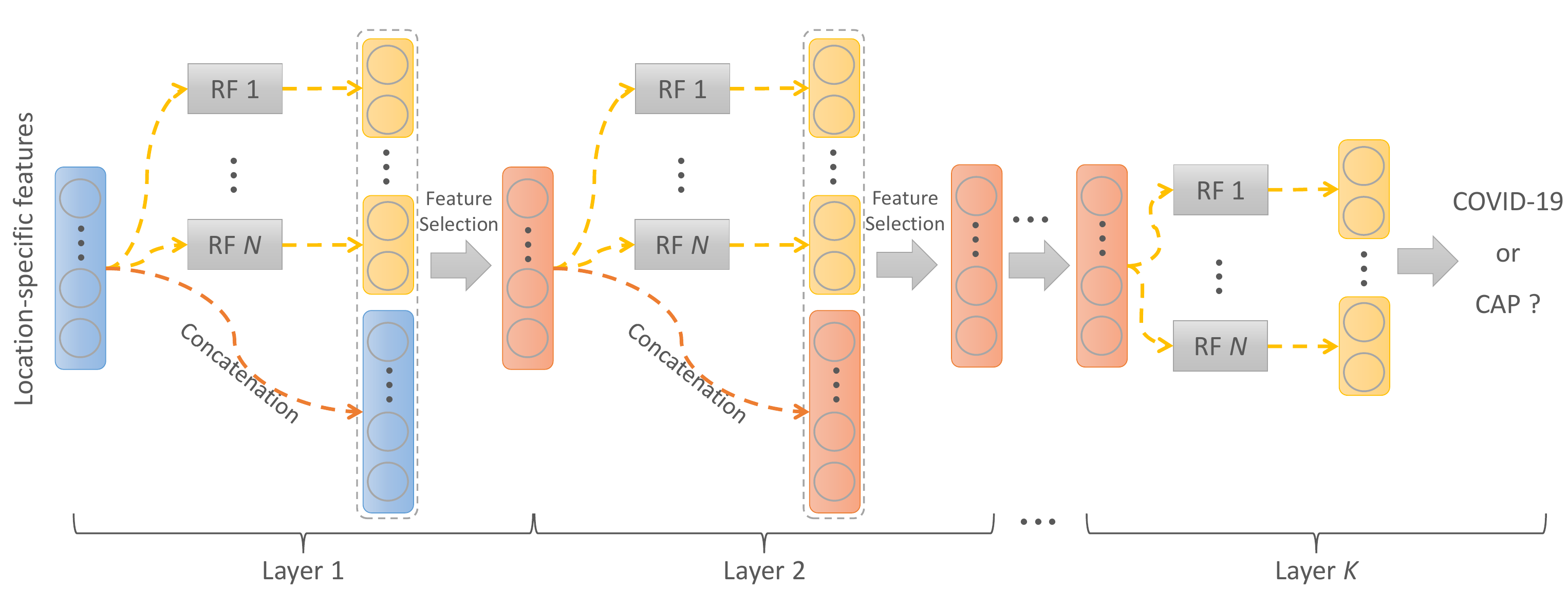}
			\caption{Overview of the proposed adaptive feature selection guided deep forest. Each layer of the proposed adaptive feature selection guided deep forest consists of $N$ random forests and an adaptive feature selection unit.}
\label{fig2}
\end{figure*}

Similar to \cite{shi2020large}, we extract the location-specific features (\ie, infection locations and spreading patterns) to represent the chest CT images for diagnosis of COVID-19. Specifically, the chest CT images are first automatically segmented into infected lung regions and lung fields bilaterally by using VB-Net\cite{shan2020lung,milletari2016v}. The infected lung regions are mainly related to mosaic sign, ground glass opacity (GGO), lesion-related signs (air bronchogram) and interlobular septal thickening. The lung fields include left lung, right lung, five lung lobes, and eighteen pulmonary segments. Then, we extract four kinds of location-specific handcrafted features, including volume, infected lesion number, histogram distribution and surface area from chest CT images. Meanwhile, we also extract the radiomics features for describing the CT images. More details are as follows.
\begin{enumerate}
\item {\bf{Volume features}:} Based on the segmented infected lung regions, we extract the total volume of infected region, and then calculate the percentage of the infected region of the whole lung. Meanwhile, according to the lung field segmentation results, we further extract the volume and percentage in each lobe and each pulmonary segment, respectively. Since there are evidence that pneumonia caused by COVID-19 more likely occurs in both right and left lungs, we also calculate the infected lesion difference as well as the percentage difference between left and right lungs.
\item {\bf{Infected lesion number}:} In comparison to CAP,  most of the COVID-19 infections encompass bilateral lungs with multifocal involvement~\cite{li2020coronavirus,chung2020ct}, and COVID-19 generally has concentrated infection lesions while CAP shows small in volume and patchy in distribution~\cite{hansell2008fleischner}. Therefore, we calculate the features of the total number of infected regions in the bilateral lungs, lung lobes, and pulmonary segments, respectively.
\item {\bf{Histogram distribution}:} The predominant chest CT findings show that bilateral and peripheral GGO and consolidation are a radiologic hallmark of COVID-19~\cite{li2020covid,bernheim2020chest}. GGO is a pattern of hazy increased lung opacity with preservation of bronchial and vascular margins, whereas consolidation is characterized by a homogeneous increase in lung parenchymal attenuation that obscures the margins of vessels and airway walls on CT images~\cite{hansell2008fleischner}. To extract the intensity distribution of the infected regions in chest CT images, we calculated the histogram features of the infected regions.
\item {\bf{Surface area}:} In previous study\cite{song2020emerging}, it has been found that COVID-19 had a predominate distribution in the posterior and peripheral lung, and the abnormalities of lung parenchyma eventually spread to the central area and bilateral upper lobes\cite{li2020coronavirus}. Therefore, we constructed the infection surface as well as the lung boundary surface. We further calculated the distance of each infection surface vertex to the nearest lung boundary surface, and categorized them into 5 ranges,~\ie, 3, 6, 9, 12 and 15 voxels. For each feature, the number of infection surface vertices within each range of distances to the lung wall is calculated. Furthermore, the percentage of infection vertex number against the number of whole infection surface vertices in each range is also considered.
\item {\bf{Radiomics features:}} Radiomics features extracted from infected lesions, including intensity features (\eg, average gray level intensity, range of gray values) and texture features (\eg, gray level co-occurrence matrix, gray-level run-length matrix, gray-level size-zone matrix, and neighborhood gray-tone difference matrix) are used in our study.
\end{enumerate}

Beside, we also adopt the age and gender into the location-specific features for the diagnosis of COVID-19. In summary, a total of 239 dimensions features are used in our study.

\subsection{Adaptive Feature Selection Guided Deep Forest}
\label{HNN}
\label{ag}
As mentioned in Section~\ref{FE}, we extract location-specific features from the chest CT images. But, simply using these features cannot adequately describe the high-level information for COVID-19 classification. In this work, we propose an adaptive feature selection guided deep forest to learn the latent high-level representation of the extracted location-specific features with adaptive task-driven feature selection process for diagnosis of COVID-19. The architecture of proposed adaptive feature selection guided deep forest is shown in Fig.~\ref{fig2}.

As shown in Fig.~\ref{fig2}, each layer of proposed adaptive feature selection guided deep forest consists of $N$ independent random forests and a feature selection unit. Here, each random forest produces a probability distribution of the COVID-19 and CAP (the yellow rectangle in Fig.~\ref{fig2}). Then, the $N$ probability distribution vector of the COVID-19 and CAP are concatenated with the input feature vector. To reduce the redundancy of the features, we further perform an adaptive feature selection operation. In particular, for each trained random forest, we can calculate the feature importance for each feature within the input feature vector. Thus, we calculate the overall feature importance $c_i$ for $i$-th feature as follows,
\begin{equation}
c_i= \frac{1}{N}\sum_{n=1}^Nc_{i,n}
\label{eq1}
\end{equation}
where $c_{i,n}$ is the feature importance for $i$-th feature in $n$-th random forest. Herein, we discard the features with low feature importance by a specific ratio based on the calculated feature importance. Hence, the feature selection and classifier training are adaptively incorporated into a unified framework. Thus, the selected feature vector as the input of the next layer. Finally, we cascade multiply layers to learn the deep discriminative feature representation for COVID-19 classification task.

%%%%%%%%%%%%%%%%% ^_^ ----------------- ^_^ %%%%%%%%%%%%%%%%%
\subsection{Implementation}
As shown in Fig.~\ref{fig2}, in the proposed adaptive feature selection guided deep forest, we employ a Xgboost~\cite{chen2016xgboost} with $20$ trees, a random forest~\cite{liaw2002classification} with $20$ trees, and two extremely randomized trees~\cite{geurts2006extremely} with $20$  and $50$ trees, respectively. We empirically set the feature discard ratio as $0.2$ in our study. In the \emph{training} stage, we feed the extracted location-specific features to the adaptive feature selection guided deep forest. The training set further is divided to $5$ subsect for $5$-fold cross-validation. In particular, each subset is sequentially selected as the validation set, while the remaining four subsets are treated as training set for model construction. Thus, the numbers of cascade layers and selected features are automatically determined by using the cross-validation strategy. Hence, the adaptive feature selection guided deep forest can adaptively train the feature selection and classification model in a task-driven manner.

In the \emph{testing} stage, we also feed the location-specific features of test subject to the trained adaptive feature selection guided deep forest model. In the last layer, each forest will produce a probability distribution $p$ for the identification of COVID-19. For each subject, we use the following equation to ensemble the predicted value for diagnose of COVID-19,
\begin{equation}
p(l=c)= \frac{1}{N}\sum_{n=1}^Np(l=c|n)
\label{eq2}
\end{equation}
where $p(l=c|n)$ is the probability of subject belongs to category $c$ (\ie, COVID-19 or CAP) that is provided by the $n$-th forest in last layer. Finally, we use the MAP criterion to obtain the label for each subject,~\ie, $argmax_cp(l=c)$
%%%%%%%%%%%%%%%%% ^_^ ----------------- ^_^ %%%%%%%%%%%%%%%%%
\section{Experiment}
\label{Experiment}
In this section, we first illustrate the competing methods and experimental settings  in our study. Then, we present experimental results achieved by different methods on the chest CT images with 1495 patients of COVID-19 and 1027 patients of CAP.
%%%%%%%%%%%%%%%%% ^_^ ----------------- ^_^ %%%%%%%%%%%%%%%%%
\subsection{Competing Methods}
In our experiments, we compare our proposed AFS-DF with the following four widely adopted machine learning methods.
\begin{enumerate}
\item {\bf{Logistic Regression (LR)}}: A Logistic regression method is employed for COVID-19 classification by using the extracted location-specific features.
\item {\bf{Support Vector Machine (SVM)}}: The extracted location-specific features are fed into the SVM classifier by using radial basis function kernel with default parameters.
\item {\bf{Random Forests (RF)}}: The random forest classifier is applied on the location-specific features for COVID-19 classification, and the number of trees in random forest is set as $500$ via cross-validation.
\item {\bf{Neural Networks (NN)}}: In this method, a fully connection neural network is employed for COVID-19 classification. Specifically, we empirically set the mini-batch size as $64$, the number of epochs as $100$, and the learning rate as $0.001$.

\end{enumerate}

\begin{table*}[!htp]
	\setlength{\belowcaptionskip}{-1pt}
	\setlength{\abovecaptionskip}{-1pt}
	\renewcommand\arraystretch{1.4}
	\centering
	\footnotesize
	\caption{Performance of COVID-19 vs. CAP classification achieved by LR, SVM, RF, NN and AFS-DF. The terms $a$ and $b$ in ``$a\pm b$'' denote the mean and standard deviation, respectively.}
	\begin{tabular*}{0.90\textwidth}{@{\extracolsep{\fill}}l cccc}
		\toprule
	    \multicolumn{1}{l}{ Method} & ACC (\%) &SEN (\%) & SPE (\%) &AUC (\%)\\
		\hline
        \multicolumn{1}{l}{\multirow{1}{*}{LR}}
		&$89.81 \pm 1.63$	&$91.64 \pm 1.69$	&$87.10 \pm 2.14$	&$95.52 \pm 1.31$  \\
		%\hline  		
		\multicolumn{1}{l}{  \multirow{1}{*}{SVM}}
		&$89.97 \pm 1.37$	&$91.51 \pm 2.04$	&$87.70 \pm 0.89$	&$95.62 \pm 0.91$ \\
		%\hline
        \multicolumn{1}{l}{  \multirow{1}{*}{RF}}
		&$89.41 \pm 1.67$	&$90.51 \pm 0.87$	& $87.85 \pm 1.01$ & $95.40 \pm 1.21$\\
		%\hline

		\multicolumn{1}{l}{ \multirow{1}{*}{NN}}
		&$89.96 \pm 1.58$	&$92.66 \pm 1.57$	&$86.04 \pm 1.97$ &$95.73 \pm 1.00$ \\
        %\hline
        %\hline
		\multicolumn{1}{l}{ \multirow{1}{*}{AFS-DF}}
		&\bm{$91.79 \pm 1.04$}	&\bm{$93.05 \pm 1.67$}& \bm{$89.95 \pm 1.28$} &\bm{$96.35 \pm 1.13$}	\\
		\bottomrule
	\end{tabular*}
	\label{tab2}
\end{table*}

\subsection{Experimental Settings}
For all extracted location-specific features from chest CT images, we first perform the normalization with center 0 and deviation 1 for each feature. Then, we employ a 5-fold cross-validation strategy to evaluate the performance of the proposed AFS-DF as well as its competitors. Specifically, all subjects are randomly partitioned into 5 subsets. Each subset is sequentially selected as the testing set, while the remaining four subsets are treated as training set for model construction, and the final classification results are the average over the five-fold cross-validations.

In order to measure the classification performance of different methods, four evaluation metrics are adopted, including classification accuracy (ACC), sensitivity (SEN), specificity (SPE) and the Area Under the receiver operating characteristic Curve (AUC)~\cite{fletcher2019clinical}. Here, the ACC, SEN and SPE are defined as,

\begin{equation}
\vspace{0ex}
\text{ACC} = \frac{\text{TP}+\text{TN}}{\text{TP}+\text{TN}+\text{FP}+\text{FN}}
\label{eq10}
\end{equation}

\begin{equation}
\vspace{0ex}
\text{SEN} = \frac{\text{TP}}{\text{TP}+\text{FN}}
\label{eq11}
\end{equation}

\begin{equation}
\vspace{0ex}
\text{SPE} = \frac{\text{TN}}{\text{TN}+\text{FP}}
\label{eq12}
\end{equation}
where TP, TN, FP and FN in Eqs.\ref{eq10}--\ref{eq12} represent True Positive, True Negative, False Positive, and False Negative, respectively.
%%%%%%%%%%%%%%%%% ^_^ ----------------- ^_^ %%%%%%%%%%%%%%%%%
\subsection{Classification Performance}
\label{Experimental Results}
We evaluate the COVID-19 vs. CAP classificationin 5-fold cross-validation on the collected chest CT images dataset. Table~\ref{tab2} shows the quantitative results (\ie, ACC, SEN, SPE and AUC) achieved by different methods.

From Table~\ref{tab2}, we can observe that our AFS-DF archives the best results in the terms of ACC, SEN, SPE and AUC. In particular, our AFS-DF method achieves the highest classification accuracy (\ie, $91.79\%$), which is better than the LR (\ie, $89.81\%$), SVM (\ie, $89.97\%$), RF (\ie, $89.41\%$) and NN (\ie, $89.53\%$). In general, the proposed AFS-DF achieves $1.98\%$, $1.82\%$, $2.38\%$ and $1.83\%$ improvements in terms of ACC over the LR, SVM, RF and NN, respectively. The possible reason for improvements is that our AFS-DF not only leverages the high-level feature representation by using the deep forest to improve the performance of COVID-19 vs. CAP, but also employs the discriminative features by using the adaptive feature selection process. The deep methods (\ie, NN and AFS-DF) show better classification accuracy in task of diagnosis of COVID-19 when compared with the conventional classifiers (\ie, LR, SVM and RF). The possible reason is that, the deep models can learn the high-level feature representative, which can boost the classification performance. It is worth noting that COVID-19 is the highly contagious disease, the higher SEN should have practically meaningful advantage for timely COVID-19 diagnosis to prevent the spread of the COVID-19. The SEN achieved by our AFS-DF for COVID-19 vs. CAP is $93.05\%$, which is better than other baseline methods. These results imply that using the adaptive feature selection guided deep forest model can improve the identification ability of COVID-19.

\begin{figure}[htp]
	\setlength{\belowcaptionskip}{-1pt}
	\setlength{\abovecaptionskip}{-1pt}
	\renewcommand\arraystretch{1.4}
	\centering
	\includegraphics[width=0.48\textwidth]{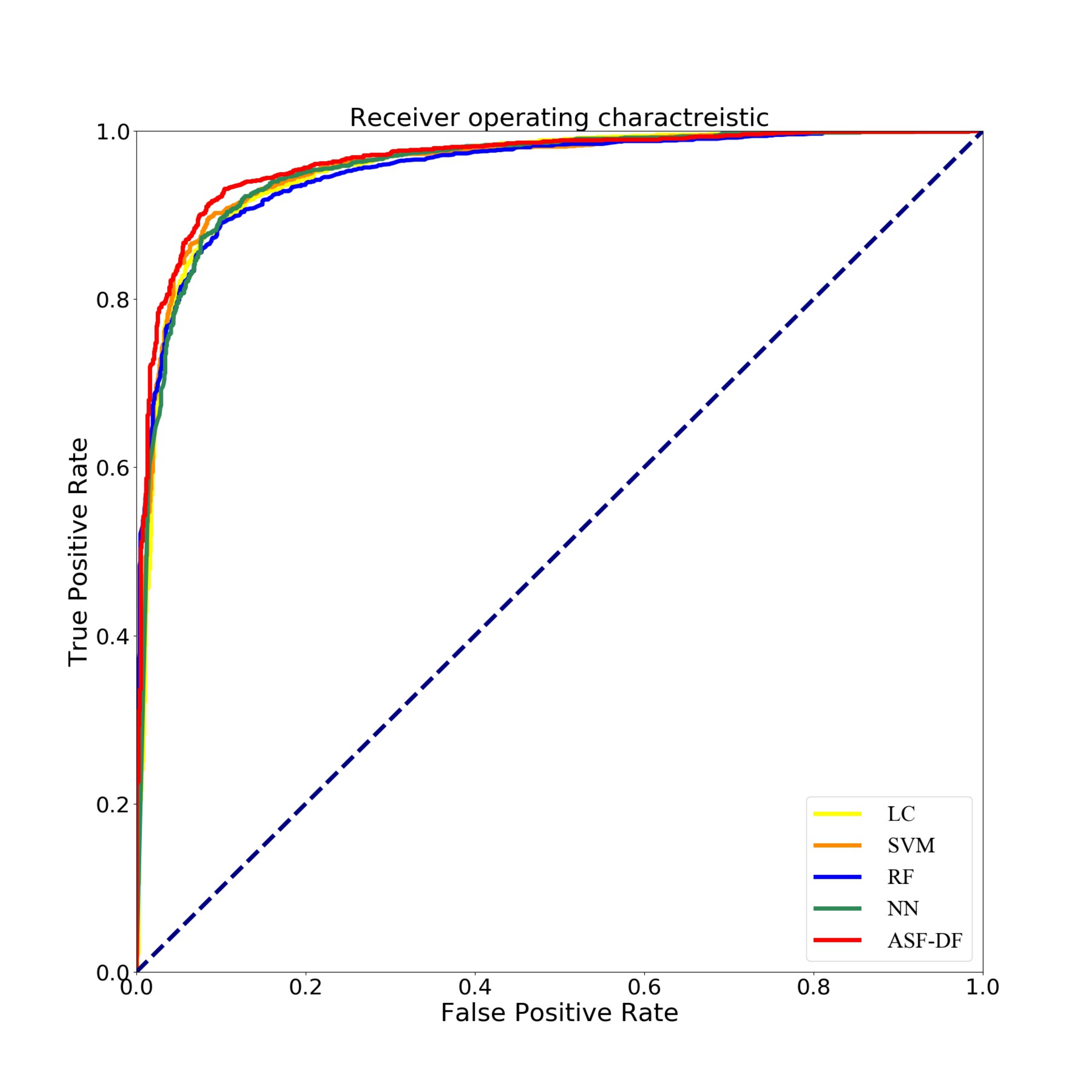}
	\caption{ROC curves achieved by LR, SVM, RF, NN and AFS-DF in COVID-19 vs. CAP classification.}
\label{fig3}
\end{figure}

\begin{figure}[htp]
	\setlength{\belowcaptionskip}{-1pt}
	\setlength{\abovecaptionskip}{-1pt}
	\renewcommand\arraystretch{1.4}
	\centering
	\includegraphics[width=0.48\textwidth]{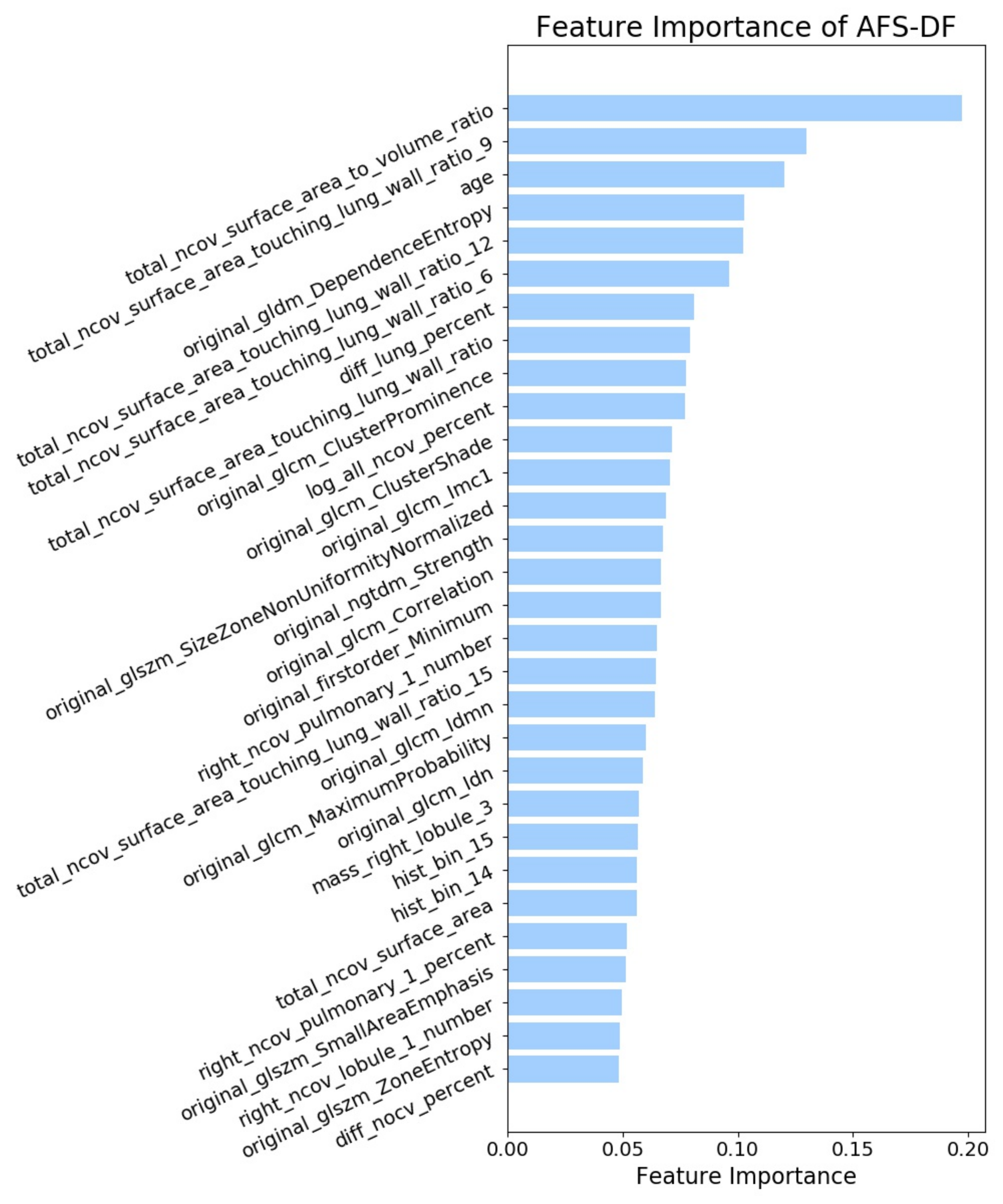}
	\caption{The top 30 important features of location-specific features in AFS-DF.}
\label{fig5}
\end{figure}

As shown in Fig~\ref{fig3}, our proposed AFS-DF method produces the best classification performance when compared with the baseline methods. These results further validate that using the high-level representation and adaptive feature selection strategy could improve the performance for COVID-19 vs. CAP classification. The statistics of feature importance of extracting location-specific features is shown in Fig.~\ref{fig5}. We calculate the sum of feature importance of location-specific features over the last layer in $5$ trained AFS-DF model. Fig.~\ref{fig5} shows that the surface area features have importance influence for COVID-19 classification.

%%%%%%%%%%%%%%%%% ^_^ ----------------- ^_^ %%%%%%%%%%%%%%%%%
\section{Discussion}
\label{Discussion}
In this section, we first compare our proposed adaptive feature selection guided deep forest with several state-of-the-art methods for COVID-19 classification. Then, we study the influence of the adaptive feature selection strategy and the selected deep features that are learned by adaptive feature selection guided deep forest. Finally, we present the limitations of this work as well as possible future research directions.

\subsection{Comparison with State-of-the-art Methods}
\begin{table}[tbp]
	\setlength{\belowcaptionskip}{-1pt}
	\setlength{\abovecaptionskip}{-1pt}
	\renewcommand\arraystretch{1.4}
	\centering
	\footnotesize
	\caption{Comparison with state-of-the-art methods for COVID-19 classification.}
	\begin{tabular*}{0.46\textwidth}{@{\extracolsep{\fill}}c cccc}
		\toprule
	    \multicolumn{1}{l}{ Method} & ACC (\%) &SEN (\%) & SPE (\%) &AUC (\%)\\
		\hline
        \multicolumn{1}{l}{\multirow{1}{*}{Wang et al.\cite{wang2020deep}}}
		&$73.1$	&$67.0$	&$74.0$ &$78.0$ \\
		%\hline  		
		\multicolumn{1}{l}{  \multirow{1}{*}{Xu et al.\cite{xu2020deep}}}
		&$86.7$	&$-$	&$-$ &$-$ \\
		%\hline
        \multicolumn{1}{l}{  \multirow{1}{*}{Shi et al.\cite{shi2020deep}}}
		&$89.0$	&$82.2$	&$82.8$ &$89.0$ \\
		%\hline
		\multicolumn{1}{l}{ \multirow{1}{*}{Tang et al.\cite{tang2020severity}}}
		&$87.5$	&$93.3$	&$74.5$ &$91.0$ \\
        %\hline
        \multicolumn{1}{l}{ \multirow{1}{*}{Shi et al.\cite{shi2020large}}}
		&$87.9$	&$90.7$	&$83.3$ &$94.2$ \\
        %\hline
		\multicolumn{1}{l}{ \multirow{1}{*}{AFS-DF}}
		&$91.79$	&$93.05$& $89.95$ & $96.35$ \\
		\bottomrule
	\end{tabular*}
	\label{tab3}
\end{table}
Since several attempts have been made for COVID-19 classification, we now compare our proposed AFS-DF with state-of-the-art methods.~\cite{wang2020deep} employs a convolutional neural networks (CNN) to extract the features of chest CT images, and then combines the decision tree and Adaboost to produce the classification result of COVID-19 vs. typical viral pneumonia (a total of 670 CT scans).~\cite{xu2020deep} leverages a ResNet~\cite{he2016deep} to predict the COVID-19, Influenza-A viral pneumonia, and healthy cases (a total of 618 CT scans). \cite{shi2020deep} extracts clinical and laboratory features and uses a logistic regression model for non-severe and severe patient classification (a total of 196 CT scans). \cite{tang2020severity} extracts quantitative features, and introduced a random forest method to assess the severity of the COVID-19 patient (a total of 176 CT scans).~\cite{shi2020large} extracts location-specific handcrafted features, and proposed an infection size-adaptive random forest for COVID-19 classification (a total of 2685 CT scans).

The results are reported in Table~\ref{tab3}. One can observe from Table~\ref{tab3} that the proposed AFS-DF shows competitive classification performance for COVID-19 patient identification. The underlying reason could be that our AFS-DF can utilize the high-level discriminative representation of the extracted features.

\subsection{Influence of Adaptive Feature Selection}
To study the effectiveness of the proposed adaptive feature selection, we compare it to two feature selection methods (\ie, Lasso and ElasticNet~\cite{zou2005regularization}) and its variant (\ie, Deep Forest (DF)). Lasso is used to select a discriminative subset of features from the feature vector by using $l_1$-norm sparsity constraint. The parameter for the sparsity constraint in this method is set as $0.001$ by using cross-validation. In ElasticNet, a $l_1$-norm is leveraged to reduce dimension of extracted features. Also, a $l_2$-norm is further introduced into the classification model to ensure the smoothness of the linear model. The parameters for the sparsity constraint and smoothness constraint in ElasticNet are set as $0.001$ and $0.1$ by using cross-validation, respectively. DF is a variant of the proposed AFS-DF, which employs the same architecture as our AFS-DF model without feature selection block. Note that this variants method has same parameters setting with AFS-DF for training and test in our study. The experimental results are reported in Table~\ref{tab5}.

As shown in Table~\ref{tab5}, the proposed AFS-DF achieves the best classification performance, when compared with Lasso, ElasticNet and DF.
In particular, compared with the feature selection methods (\ie, Lasso and ElasticNet), the proposed AFS-DF achieves better performance by using the high-level feature representation. Meanwhile, by using the adaptive feature selection operation, the AFS-DF achieve a better performance when compared with DF. These results imply the effectiveness of the proposed AFS-DF.

\begin{table}[tbp]
	\setlength{\belowcaptionskip}{-1pt}
	\setlength{\abovecaptionskip}{-1pt}
	\renewcommand\arraystretch{1.4}
	\centering
	\footnotesize
	\caption{Performance of COVID-19 vs. CAP classification by using Lasso, ElasticNet, DF and AFS-DF. The terms $a$ and $b$ in ``$a\pm b$'' denote the mean and standard deviation, respectively.}
	\begin{tabular*}{0.50\textwidth}{@{\extracolsep{\fill}}l ccc}
		\toprule
	    \multicolumn{1}{l}{ Method} & ACC (\%) &SEN (\%) & SPE (\%)\\
		\hline
        \multicolumn{1}{l}{\multirow{1}{*}{Lasso}}
        &$90.33 \pm 0.87$	&$92.51 \pm 1.25$	&$87.14 \pm 0.53$\\
        %\hline
        \multicolumn{1}{l}{\multirow{1}{*}{ElasticNet}}
		&$90.25 \pm 1.26$	&$92.71 \pm 1.41$	&$86.68 \pm 1.35$\\
		%\hline
         \multicolumn{1}{l}{ \multirow{1}{*}{DF}}
		&$91.16 \pm 1.34$	&$92.85 \pm 1.54$ & $88.65 \pm 1.59$	\\
        %\hline
		\multicolumn{1}{l}{ \multirow{1}{*}{AFS-DF}}
		&\bm{$91.79 \pm 1.04$}	&\bm{$93.05 \pm 1.67$}& \bm{$89.95 \pm 1.28$}\\
		\bottomrule
	\end{tabular*}
	\label{tab5}
\end{table}

\subsection{Influence of Features}
\label{ihf}
To evaluate the effectiveness of the selected deep features (\eg, the features used in the last layer of AFS-DF) for COVID-19 vs. CAP classification, we further develop three methods based on LR, SVM and RF by using the selected deep features (\ie, AFSDF-LR, AFSDF-SVM and AFSDF-RF). We evaluate these six methods for COVID-19 vs. CAP classification, with the results reported in Table~\ref{tab4}.

As can be seen from Table~\ref{tab4}, the proposed AFSDF-LR, AFSDF-SVM and AFSDF-RF outperform their counterparts (\ie, LR, SVM and RF) in most of evaluation metrics. Of note, the proposed methods consistently achieve better results in terms of ACC and SEN. For example, AFSDF-LR, AFSDF-SVM and AFSDF-RF achieve $1.38\%$, $1.15\%$ and $1.11\%$ improvement over LC, SVM and RF in terms of ACC for COVID-19 vs. CAP classification, respectively. Compared with LC, SVM and RF, the AFSDF-LR, AFSDF-SVM and AFSDF-RF also show improvement in terms of SEN for COVID-19 vs. CAP classification, respectively. The possible reason is that, with the selected deep features by using AFS-DF, the features include the high-level and discriminative information. Hence, the conventional machine learning methods (\ie, LR, SVM and RF) can use these selected deep features to improve the performance of COVID-19 classification task.

Beside, In Fig~\ref{fig4}, we plot the original location-specific features, the features of the last layer in DF, and the features of the last layer in AFS-DF, which perform dimensionality reduction by using \emph{t}-SNE~\cite{maaten2008visualizing}. As shown in Fig~\ref{fig4}, our AFS-DF produces more discriminative features for COVID-19 vs. CAP classification.

\begin{table}[tbp]
	\setlength{\belowcaptionskip}{-1pt}
	\setlength{\abovecaptionskip}{-1pt}
	\renewcommand\arraystretch{1.4}
	\centering
	\footnotesize
	\caption{Performance of COVID-19 vs. CAP classification by using LR, AFSDF-LR, SVM, AFSDF-SVM, RF and AFSDF-RF. The terms $a$ and $b$ in ``$a\pm b$'' denote the mean and standard deviation, respectively.}
	\begin{tabular*}{0.50\textwidth}{@{\extracolsep{\fill}}l ccc}
		\toprule
	    \multicolumn{1}{l}{ Method} & ACC (\%) &SEN (\%) & SPE (\%)\\
		\hline
		\multicolumn{1}{l}{\multirow{1}{*}{LR}}
		&$89.81 \pm 1.63$	&$91.64 \pm 1.69$	&$87.10 \pm 2.14$ \\
		%\hline
        \multicolumn{1}{l}{\multirow{1}{*}{AFSDF-LR}}
		&$91.19 \pm 1.11$	&$92.52 \pm 1.51$	&$89.08 \pm 2.00$ \\
		%\hline  				
		\multicolumn{1}{l}{  \multirow{1}{*}{SVM}}
		&$89.97 \pm 1.37$	&$91.51 \pm 2.04$	&$87.70 \pm 0.89$\\
        %\hline
        \multicolumn{1}{l}{\multirow{1}{*}{AFSDF-SVM}}
		&$91.12 \pm 1.50$	&$92.31 \pm 1.82$	&$89.38 \pm 1.52$ \\		
		%\hline
        \multicolumn{1}{l}{  \multirow{1}{*}{RF}}
		&$89.41 \pm 1.67$	&$90.51 \pm 0.87$	& $87.85 \pm 1.01$\\
		%\hline
        \multicolumn{1}{l}{  \multirow{1}{*}{AFSDF-RF}}
		&$90.52 \pm 1.21$	&$92.51 \pm 1.27$	& $87.57 \pm 2.50$\\
		%\hline
		\bottomrule
	\end{tabular*}
	\label{tab4}
\end{table}

\begin{figure}[htp]
	\setlength{\belowcaptionskip}{-1pt}
	\setlength{\abovecaptionskip}{-1pt}
	\renewcommand\arraystretch{1.4}
	\centering
	\includegraphics[width=0.40\textwidth]{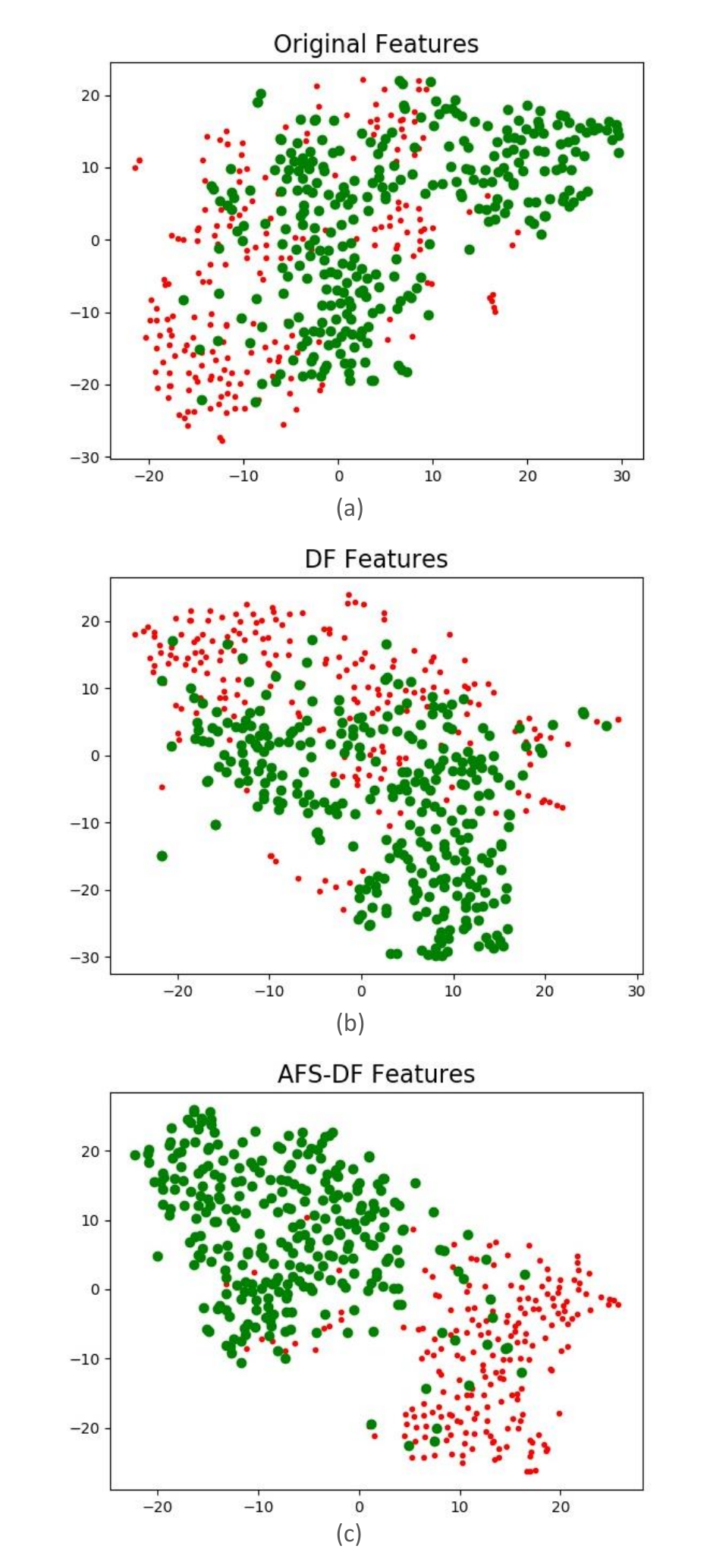}
	\caption{Visual illustration of original location-specific features (a), the features of the last layer in DF (b), and the features of the last layer in AFS-DF (c).}
\label{fig4}
\end{figure}

%%%%%%%%%%%%%%%%% ^_^ ----------------- ^_^ %%%%%%%%%%%%%%%%%
\subsection{Limitations and Future Work}
There are still several limitations in the current study. \emph{First}, the adaptive feature selection guided deep forest is only validated on COVID-19 vs. CAP classification task. In the future, we plan to perform our proposed method on other COVID-19 classification tasks (\eg, COVID-19 vs. normal, severe patients vs. non-severe patients, \etc). \emph{Second}, we extract the handcraft features by using prior knowledge in current work; in future, the features learned by deep learning method are expected to leverage our proposed method for further performance improvement.

%%%%%%%%%%%%%%%%% ^_^ ----------------- ^_^ %%%%%%%%%%%%%%%%%
\section{Conclusion}
\label{Conclusion}
In this paper, we propose an adaptive feature selection guided deep forest for COVID-19 vs. CAP classification by using the chest CT images. Specifically, the AFS-DF uses the deep forest to learn the high-level representation based on the location-specific features. Meanwhile, an adaptive feature selection operation is employed to reduce the redundancy of features based on the trained forest. Experimental results on the collected COVID-19 dataset with 1495 COVID-19 cases and 1027 CAP cases show that our proposed AFS-DF approach can achieve superior performance on COVID-19 classification with chest CT images in comparison with several existing methods.

\footnotesize
\bibliographystyle{IEEEtran}
\bibliography{reference}

\end{document}